\documentclass[prl,twocolumn]{revtex4}
\usepackage{epsfig}
\usepackage{graphics}
\begin{document}
\title{Quantum Spin Dynamics of Spin-1 Bose Gas}
\author{Roberto B. Diener and Tin-Lun Ho}
\affiliation{Department of Physics, The Ohio State University, Columbus, OH 43210, USA}
\date{\today}

\begin{abstract}
We show that the quantum evolution of a spin-1 Bose gas with nearly all bosons initially in the $F_z = 0$ state has a ``quantum carpet" {\em spin-time} structure with self-similar properties. The system continuously evolves into ``multi-peaked" Schr\"odinger cat like states, returning occasionally  to coherent structures, which leads to large number fluctuations (as seen in recent experiments). 
The self similar behavior allows one to reveal the quantum evolution as a set of peaks in the number probability distribution of a spin component at times {\em much shorter} than the quantum revival time.  We also show that these features survive small number fluctuations among spin components up to a few percent. 
\end{abstract}

\maketitle

Since the discovery of Bose-Einstein condensation in alkali atomic gases, there have
been increasing activities in Bose gases with non-zero spin $F$.  
Compared to scalar Bose gases, these systems have a greater variety of ground states and  macroscopic quantum phenomena.  Bosons with spin also show great promise for quantum computation and quantum information processing, since their spin-dependent Hamiltonians have been derived and are completely specified by the scattering lengths in different angular momentum channels~\cite{Hospinor}. Despite the simple appearance of these Hamiltonians, their dynamics is surprisingly rich.  Understanding spin dynamics, and in particular {\em quantum} spin dynamics, is important not only for theoretical reasons, but also for engineering and controlling the time evolution of spin carrying bosons. 

Recently, the spin dynamics of $F=1$ and $F=2$ Bose gases of $^{87}$Rb have been experimentally studied~\cite{experiments}. 
For $F=1$ $^{87}$Rb bosons, if initially all bosons are in the $F_{z}=0$ level,  the number of bosons in this level at later times (when averaged over many runs)  exhibits a ``damped" oscillatory behavior {\em with very large fluctuations}, whereas the fluctuations in the number difference between 
$F_{z}=1$ and $F_z={-1}$ spin states is almost  zero (due to spin conservation).  
At present, the theoretical explanation for this oscillatory behavior is based on mean field spin dynamics within the so-called single mode approximation (SMA), with an averaging over the initial condition to simulate 
the slightly different initial conditions in different experimental runs.  (SMA means all spin states have the same spatial density profile). 
The large fluctuations are due to the sensitive dependence of the semiclassical evolution on the initial conditions~\cite{Pu}.
Mean field spin dynamics assumes that the system remains an {\em un-fragmented} condensate at all times. 
This assumption is considered satisfactory because the mean field calculations seem to account for the observed damped oscillation in the $F_{z}=0$ component. There is, however, no direct experimental verification of it.  

On the other hand, it is known that quantum dynamics also exhibits a ``damped" oscillation with large fluctuations in the $F_{z}=0$ component~\cite{Law}, although this oscillation 
is in fact non-dissipative with a period $t_o$. In current experiments this quantum revival time is longer than the duration of the experiments and is thus not observable.  In this letter we present a method for distinguishing between the semiclassical and quantum dynamics in time scales much shorter than $t_o$.  We shall show that hidden in the quantum spin dynamics of spin-1 Bose gases lies a set of beautiful scaling relations in the noise statistics. Not only are they dramatic demonstrations of quantum behavior, but also practical diagnostics that allow us to detect the existence of quantum revivals at times smaller than the full revival period.


\begin{figure*}
\centering \epsfig{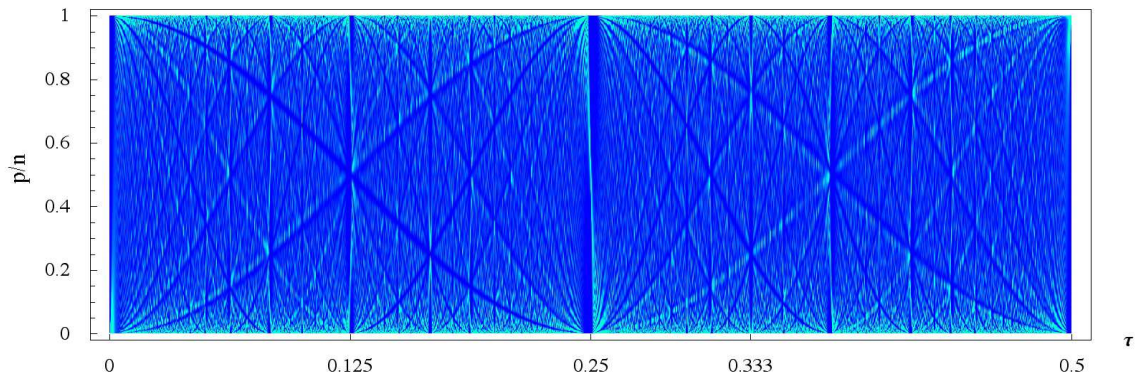}
\caption{Probability of finding $p$ bosons in $F_{z}=+1$ state as a function of time $\tau$ for 1000 bosons.  Light (dark) colors are regions of large (small) probability.  At $\tau=0$ ($\tau =1/2$), the probability is sharply peaked at $p=0$ ($p = n$).   In between, the distribution of probabilities has a fractal behavior.} \label{carpet.fig}
\end{figure*}

{\bf (A) Mean field spin dynamics:}
Within SMA, the Hamiltonian of  $N$ spin-1 bosons in zero magnetic field is\cite{Hospinor,note on c}
$H_{o}=c {\bf S}^2, \,
{\bf S} =\sum_{\mu\nu} a^{\dagger}_{\mu}{\bf S}_{\mu\nu}^{} a^{}_{\nu}$,
where $a^{\dagger}_{\mu}$ creates a boson in the state $F_{z} = \mu$ and ${\bf S}_{\mu\nu}$ is the spin-1 matrix.  We can also write ${\bf S}^2 = (\hat{N}_1-\hat{N}_{-1})^2+(2\hat{N}_0-1)(\hat{N}_1+\hat{N}_{-1})+(2 a_1^\dagger  a_{-1}^\dagger a_0^2+ {\rm h.c.})$, where $\hat{N}_{\mu} = a^{\dagger}_{\mu} a^{}_{\mu}$ is the number of bosons in state 
$\mu$.  The Hilbert space is spanned by the states $|N^{}_{1}, N^{}_{0}, N_{-1}^{}\rangle$ which have $N_{\mu}^{}$ bosons in state $\mu$.  In the presence of a magnetic field ${\bf B} = B\hat{\bf z}$, there are linear and quadratic Zeeman effects given by $H_{1} = -p (\hat{N}_{1}-\hat{N}_{-1})$ and $H_{2} = \alpha (\hat{N}_{1}^2  + \hat{N}_{-1}^2)$, where $p$ and $\alpha$ are proportional to $B$ and $B^2$ respectively.  Due to the conservation of the total spin $S_z = M$, the dynamics is independent of $p$.  Since we are interested in low field phenomena, we will consider the case $\alpha=0$.  

The mean field approach assumes that the system is in a coherent state 
$|\Psi(t)\rangle = (\sum_{\mu}\zeta_{\mu}(t) a^{\dagger}_{\mu})^{N}|0\rangle/\sqrt{N!}$, where $\zeta_{\mu}(t)$ is a normalized vector. The single particle density matrix is
$\rho_{\mu \nu}^{} = \langle \Psi(t) |a^{\dagger}_{\mu} a^{}_{\nu} | \Psi(t) \rangle = 
N\zeta^{\ast}_{\mu}(t) \zeta^{}_{\nu}(t)$, which has only one eigenvalue.  
The occupation number of state $\mu$ is $N_{\mu}(t) = |\Phi_{\mu}^{}(t)|^2$, where $\Phi_{\mu}^{}(t) = \sqrt{N}\zeta_{\mu}^{}(t)$ is the condensate wavefunction which evolves according to the mean field equation
$ i\hbar\partial_{t}^{} \Phi_{\mu}^{}(t) = 2 c \langle {\bf S}(t)\rangle \cdot {\bf S}_{\mu\nu}^{} \Phi_{\nu}^{}(t)$, 
and $\langle {\bf S}(t)\rangle = \sum_{\alpha\beta}^{} \Phi^{\ast}_{\alpha}(t) {\bf S}^{}_{\alpha\beta}\Phi^{}_{\beta}(t)$. If initially all the bosons are in the $\mu = 0$ state as in ref.  \cite{experiments}, (i.e. $\Phi_{\mu}(0) = \sqrt{N}\delta_{\mu,0}$) the system does not evolve, since the mean field equations  imply that such a state is stationary.  In practice it is assumed that in each run of the experiment there is a small but non-zero $\Phi_{\pm 1}(0)$.  The experimental results are to be compared to the average of such runs, which yields a decay of the oscillations in $N_{0}(t)$ with large fluctuations.

{\bf (B) Quantum  spin dynamics:}  In contrast, in the quantum dynamics the state $|0,N,0\rangle$ is not stationary, due to the ``conversion" term $a^{\dagger}_{1} a^{\dagger}_{-1} a^{2}_{0}$ in ${\bf S}^2$.  In the following, we shall focus on the evolution of this state, since it is related to the experimental situation, and is the simplest case that captures most of the intricacy and richness of the quantum evolution. For simplicity, we shall consider a system with an even number of bosons, $N=2n$, where $n$ is an integer.  Since $H_{o}$ is rotationally invariant, 
it only couples  the initial state $|0,N,0\rangle$ to states $|p\rangle \equiv |p, N-2p, p\rangle$, where $p$ is the number of bosons in state $\mu=\pm 1$, $(0\leq p\leq n)$.   We can then expand the quantum state at time $t$ as 
$
|\Psi (t)\rangle = \sum_{p=0}^{n} \Psi (p,t)|p\rangle, \,\,\,\,\,\,\, \Psi (p,t) = \langle p |e^{-iH_{o}t /\hbar} |0\rangle, 
$
with initial condition $\Psi(p,0) = \delta_{p,0}$. 
For operators $\hat{O}$ that are diagonal in number space, their expectation values are 
$\langle \hat{O}(t) \rangle = \sum_{p} O(p) |\Psi (p,t) |^2$. For $N=2n>>1$, it is convenient to write 
\begin{equation}
p/n = (\cos \theta + 1)/2, \,\,\,\,\,\,\, 0\leq \theta \leq \pi, 
\label{theta} \end{equation}
$\Psi(p,t)\equiv \Psi(\theta, t)$, so that 
$\langle \hat{O}(t) \rangle = \int^{\pi}_{0} {\rm d}\theta \,O(\theta) \,|f(\theta,t)|^2$,
 where $f(\theta,t)= \sqrt{(n/2) {\rm sin}\theta}\, \Psi(\theta, t)$. 
The key features of the quantum dynamics is contained in the behavior of $f(\theta)$. The system will be either a number (i.e. Fock) state or a coherent state if the probability distribution ${\cal P} (\theta, t) \equiv |f(\theta, t)|^2$ is a delta-function or a sharply peaked function of width $\sqrt{n}$, with a number fluctuation $(\Delta N_{0})^{2} = \langle ( \hat{N}_{o} -  \langle \hat{N}_{o}\rangle)^2 \rangle$
ranging from 0 to $n$. If, however,  $f(\theta, t)$ is a 
sum of several sharply peaked distributions, the system is in a Schr\"odinger cat like state, with $(\Delta N_{0})^{2}$ of order $n^2$. 

{\bf (B1) Quantum Carpet:}  Since $H_{o}= c {\bf S}^2$,
 \begin{equation}\label{time evolution}
\Psi(p,\tau)= \sum_{S} \langle p| S,0\rangle \langle S,0|0\rangle  e^{-i \pi S (S+1)\tau}, 
\label{psipt}\end{equation}
where $\tau = t/t_{o}$, $t_{o} = \pi \hbar/c$ (independent of number of particles),  $|S, M\rangle$ is the state with total angular momentum $S$, and $S_{z} = M$. Since $N$ is even, Bose statistics implies that $S$ is also even, and we shall write $S=2k$, where $0\leq k\leq n$.  Since $\langle p | S, 0 \rangle$ is real (see Appendix), it is clear from eq.(\ref{psipt}) 
\begin{equation}
f(\theta, \tau) = f(\theta, \tau+1), \,\,\,\,\,\,\,\,\, f(\theta, -\tau) = f(\theta, \tau)^{\ast}. 
\label{simple} \end{equation}

Using the exact expressions for $\langle p|2k, 0\rangle$ derived in the Appendix, we can calculate the probability $|\Psi(p,\tau)|^2$ exactly for any number of particles. In figure 1, we display  $|\Psi(p,\tau)|^2$  in a ``spin-time" plot for a system of 1000 particles,  where light color means high probability. 
It is a highly intricate yet orderly self similar  pattern, analogous to the so-called ``quantum carpet" that displays the motion of a quantum mechanical particle in a one dimensional box\cite{Berry}.  At $\tau=0, 1/2,$ and $1/4$,  $|\Psi(p,\tau)|^2$ consists of a single sharp peak at $p=0$,  $n$, and $n/2$ respectively. They are therefore either a Fock state or a coherent state. At any other time, however,  $|\Psi(p,\tau)|^2$ has many different peaks and the system is therefore Schr\"odinger cat-like. For example, at time $\tau = 1/8$ and $1/3$ one can see from  fig.\ref{carpet.fig} that the distribution consists of two sharp peaks.

To understand these intricate features we need to analyze the spinor wavefunction and their initial probability amplitudes $c_k =  \langle 2k, 0|0\rangle$.  In the appendix we show that 
 for $1\ll k \ll n, p, (n-p)$, the spinor wavefunction can be approximated by 
(up to $O(1/n)$)~\cite{not valid for theta = 0}, 
\begin{equation} 
\langle p|2k,0\rangle \quad \approx \sqrt{4\over \pi n}{(-1)^{p+k}\over \sqrt{\sin \theta}}\, \cos \left [ (k+{1\over 4}) \theta \right ]; 
\label{p2k0}
\end{equation}
and that to the same accuracy, we also have 
\begin{equation}
c_k \approx \sqrt{4k+1\over 2n}\, e^{-k^2/(2n)} \approx A e^{(k-k_{o})^2/\sigma^2}
\label{02kapp}\end{equation} 
where $k_{o} \approx 0.8 \sqrt{n}$, $\sigma \approx \sqrt{n}$, and $A$ is a constant. Thus the initial state populates angular momentum states up to $S$ of order $\sqrt{n}$ and we are allowed to use (\ref{p2k0}).
We then have 
\begin{equation}
f(\theta, \tau) =    \sum_k  \sqrt{2\over \pi}c_k\,  e^{-i2\pi k(2k+1)\tau+i\pi(p+k)} \cos \left [(k+{1\over 4})\theta \right ]. 
\label{sum} \end{equation} 

It proves useful to extend $f$ outside the physical interval $[0, \pi]$ to all $\theta$, so that $f(\theta, \tau) = f(\theta + 8\pi, \tau)$. 
Eq.(\ref{sum}) implies that $f(\theta, 0)$ is a sequence of sharp peaks at 
$(2\ell + 1)\pi$ (where $\ell$ is an integer)  such that $f(\pi, 0)=  - f(3\pi, 0)= - f(5\pi, 0) = + f(7\pi, 0)$.  It is straightforward to show from eq.(\ref{sum}) that 
\begin{eqnarray}
f(\theta, \tau + \frac{1}{4}) &=& \frac{e^{-i\pi/8}}{2}\sum_{\pm}\left[ f(\theta \pm  \frac{\pi}{2}, \tau)  + i f(\theta \pm \frac{3\pi}{2}, \tau)\right]\nonumber \\
f(\theta, \tau + {1\over 2}) &=& (f(\theta - \pi, \tau)+ f(\theta + \pi, \tau) )/\sqrt{2}. \label{12}
\end{eqnarray}
Finally, eq.(\ref{sum}) also implies (shown in next section) that if $\theta/(4\tau +1) \ll \sqrt{n}$
\begin{equation}
f(\theta, \tau) = \sqrt{   \frac{-i}{4\tau +1}   } e^{i\zeta(\tau)}
f\left( \frac{\theta}{4\tau + 1}, \frac{\tau}{4\tau +1}\right)
\label{scale}\end{equation}
where $\zeta(\tau) = \frac{i\pi}{4\tau + 1}( \tau^2 + (\theta/2\pi)^2)$.

We are now ready to study fig.1. Consider the dense set of rational times $\tau=P/Q$, where $P$ and $Q$ are integers relatively prime to each other, and note that 
 
\noindent ${\bf (a)}$: Eq.(\ref{simple}) implies any rational $\tau = P/Q>1/2$ can be ``mapped" onto $P'/Q < 1/2$ with  $P'<P$.  It is thus sufficient to consider rationals in $[0, 1/2]$. 
  
\noindent   ${\bf (b)}$: Eqs. (\ref{simple}) and (\ref{scale}) imply $P/Q  \, \epsilon\, (0, 1/2)$ 
(except for 1/4) can be mapped onto $\tau' = P/Q'$ with $Q'= (4P-Q)$ such that $|Q'|<Q$. 

\noindent   ${\bf (c)}$: Since $f(\theta, 0)$ is sharply peaked at $(2\ell +1) \pi$, Eq.(\ref{12}) implies $f(\theta, 1/2)$ peaks at  $2\ell \pi$ and $f(\theta, 1/4)$ peaks at  $(\ell + 1/2)\pi$.  The latter is due to the fact that $f(\theta, 1/4)$ is a sum of $f(\theta, 0)$'s moved forward and backward by $\pi/2$. 

Since both ${\bf (a)}$ and ${\bf (b)}$ reduce $P$ and $Q$, their repeated application
can reduce the distribution at $\tau = P/Q$ to that at $\tau_F = 1/Q_F = 1, 1/2$, or $1/4$, depending on whether $Q$ is odd, of the form $4m-2$ or $4m$ respectively.  Using the fact that after one application of eq.(\ref{scale}) the $\theta$ variable gets scaled by a factor $Q'/Q$, successive application of ${\bf (a)}$ and $(\bf b)$ to the final stage will scale  $\theta$ by a factor  $Q_F/Q$.  
This leads to the following ``Rules" for the peaks $\theta^{\ast}$ of the  distribution function.  

\noindent {\bf (1)} if $Q = 2m-1$, $Q_{F}/Q=1/Q$, then $\theta^{\ast} = \pi(2\ell+1)/Q$,

\noindent {\bf (2)} if $Q = 4m-2$, $Q_{F}/Q=2/Q$,  then  $\theta^{\ast} = 4\ell \pi /Q$,

\noindent {\bf (3)}  if $Q = 4m$, $Q_{F}/Q=4/Q$,  then  $\theta^{\ast} = (2+4\ell)\pi/Q $; 

\noindent where  $\ell = 0, ..., m-1$. In each case there are $m$ peaks in the physical interval $[0,\pi]$.  

As an example of the application of these rules,  we list the peak locations of the probability $|\Psi(p,P/Q)|^2$:  

\begin{centering}
\begin{tabular}{|c|c|c|c|c|c|} 
\hline
Q & Case & $m$ & $\theta^{\ast}$ \hspace{0.5in}& $\Delta \theta^{\ast}$ &$p^{\ast}/n$ \hspace{0.5in}\\ \hline
2& ${\bf (2)}$ & 1 &0 &- &1/2 \\ \hline
3& ${\bf (1)} $ &2 & $\pi/3, \,\,\, \pi $&$2\pi/3$  & 1/4, 0 \\ \hline
4& ${\bf (3)}$ & 1 & $\pi/2 $ &- & 1/2 \\ \hline
5 & ${\bf (1)}$ & 3 & $\pi/5, \,\,3\pi/5,\,\,  \pi $ &$2\pi/5$  & eq.(\ref{theta}) \\ \hline
20 & ${\bf (3)}$ & 5 &$k \pi/10, \,\, k=1,3,5,7,9$ &$\pi/5$ & eq.(\ref{theta}) \\ \hline
\end{tabular}
\end{centering}

\begin{figure}[t]
\centering \epsfig{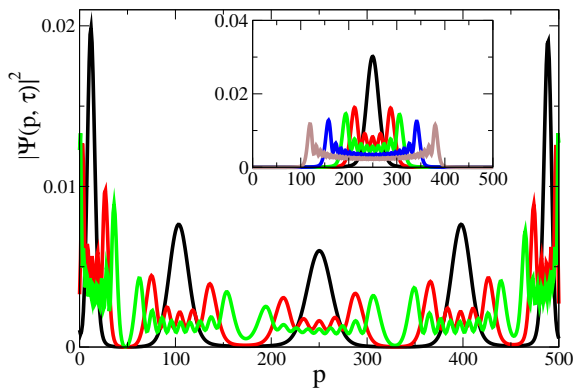}
\caption{Probability of finding $p$ bosons in the $F_{z}=1$ state at $\tau = 1/20$ for $N=2n=1000$.  The black, red, and green lines correspond to the initial state $|q,2(n-q),q\rangle$ with $q=0, 4, 8$ resp.  The inset shows the probabilities for $\tau = 1/4$ for $q = 0, 4, 8, 20, 40$ (black, red, green, blue and black lines, resp.).  }\label{figure.fig}
\end{figure}

\noindent The case $\tau=1/20$ is shown (as the black curve) in the main part of fig. \ref{figure.fig}. 

The intricate behavior in fig.\ref{carpet.fig} is due to the fact that as $\tau$ increases, it sweeps through a dense set of rationals where $Q$ changes in a chaotic fashion. The important point, however, is that {\em even when the quantum revival time is longer than the duration of experiment,  its presence 
is
reflected as a sequence of distinct peaks in the probability distribution at shorter times}.
The observation of these peaks in the probability distribution would be a clear demonstration of quantum evolution.

\vspace{0.1in}

{\bf (C) Derivation of eq.(\ref{scale}):}  Recall that the elliptic theta-function 
$\Theta_{3}(v,\tau) = \sum_{k} (e^{i\pi\tau})^{k^2} e^{2i\pi kv}$ obeys
\begin{equation}
\Theta_{3}(v,\tau) = \Theta_{3}(v, \tau + 2) = \Theta(v+1, \tau), 
\label{propsTheta}\end{equation}
and satisfies  the reciprocal relation\cite{book}
\begin{equation}
\Theta_{3}(v,\tau) = \sqrt{\frac{i}{\tau}} e^{-i\pi v^2/\tau}
\Theta_{3}\left( \frac{v}{\tau}, - \frac{1}{\tau}    \right).
\label{reci} \end{equation}
It is then straightforward to show that eq.(\ref{sum}) is 
\begin{eqnarray}
 f(\theta, \tau)  = & B(-1)^p \sum_{\pm} e^{\pm i \theta/4} \hspace{1.0in} \nonumber \\
 & \times 
  \Theta_3\left( 
\pm  \frac{\theta}{2\pi} - \tau - i k_{o}\epsilon,  -(4\tau +1)+i \epsilon \right)
 \label{thetafunctions}
\end{eqnarray}
where $\epsilon = 1/(\pi \sigma^2)$ and $B$ is a constant .  To obtain the scaling relation (\ref{scale}) we apply (\ref{reci}) to both theta functions in eq.(\ref{thetafunctions}), neglecting the imaginary parts in the exponential parts. This is valid as long as $\theta k_0 \epsilon \ll (4\tau +1)$. 
%
 If we further use (\ref{propsTheta}) to subtract $2$ from the second variable of the Theta function and use that $\Theta(-v, \tau) = \Theta(v, \tau)$,  we obtain (\ref{scale}). 
The restriction on the imaginary parts shows that the maximum value of $Q$ for which the rules can be applied is of the order $\sqrt{n}$.

{\bf (D) Fluctuations in initial states:} 
Since the semiclassical evolution of states close to $|0,N,0\rangle$  is highly sensitive to initial condition, we need to consider how the quantum evolution is affected by small fluctuations in particle numbers in various spin components.  The state of interest is now  
$|\Psi(t)\rangle = \sum_{p}\Psi(p,t) |p; M\rangle$, where $ |p; M\rangle\equiv |p+M,  N-M -2p  , p\rangle$
and $M$ is the magnetization. The initial condition is $|\Psi(0)\rangle =  |q; M\rangle$ with  $q, M<<N$, and we are interested in  $\Psi(p, \tau) = \sum_{k} \langle p; M|S, M\rangle\langle S, M|q; M\rangle
e^{-i \pi S(S+1)\tau} $.   
We present here the numerical results and describe the phenomena but we shall present the details elsewhere~\cite{longer paper}.

Let us consider the case $M=0$, $n_{1}=n_{-1} =q$.  When $q\neq 0$, each of the $m$ peaks in the original probability distribution $|f(\theta, P/Q)|^2$ (separated by $\Delta \theta^\ast$ in $\theta$-space) is  split into a group of $q+1$-subpeaks centered around the original peak at $\theta^{\ast}$, such that the outermost two peaks have the highest probability. The range of the splitting, i.e. the distance between the two outermost peaks,  is  
$\delta \theta_q = 4 \sqrt{q/n}$~\cite{longer paper}.  In order for the original peaks to remain as separate groups, one needs $\delta \theta_q < \Delta \theta^{\ast}$.  The critical fluctuation $(q/n)_c$ beyond which different groups begin to interfere is $(q/n)_c = (\Delta \theta^{\ast}/4)^2$.

This phenomenon is demonstrated in fig.\ref{figure.fig}, where we have plotted $|\Psi(p, \tau)|^2$ for a system of 1000  bosons with $\tau=1/20$  for different number fluctuations $q=0, 4, 8$, corresponding to $q/n= 0, 0.008, 0.016$ respectively.  In this case $Q=20$, $\Delta \theta^\ast = \pi/5$, and we have  $(q/n)_c = 0.025$.  For $q=4$, $q/n<(q/n)_c$, we see that while each original peak is split into five subpeaks, each group of peaks is still separated from the rest.  
Another example of the effect of $q$ on the original peaks is shown in the inset for  the probabilities at $\tau = 1/4$ for $q=0, 4, 8, 20$, and $40$, with $q/n$ ranging from $0$ to $8\%$.  Given that different values of $q$ are realized in different runs, experiments sample the average of these curves.  What is remarkable is that for the case of $\tau=1/4$, the original single peak structure in the probability distribution is still reflected in the averaging process for $q/n$ as high as $8\%$. 
These behavior also applies to much larger systems, as well as states with $M\neq 0$.  

We have therefore demonstrated that the peak structure in the probability distribution at rational times, which is the hallmark of quantum evolution, survives number fluctuations in different spin components as long as they are no more than a few percent. 

{\bf Appendix:} 
The fully polarized spin states $|S=2k,2k\rangle$ can be expressed as~\cite{HoYip} $a_1^{\dagger 2k}\,\Theta^{\dagger n-k} |\emptyset \rangle/\sqrt{f(n; k)}$ where the normalization constant is $
f(n;k) = (2k)!^2\, (n-k)!\, (2n+2k+1)! /[(n+k)! (4k+1)!]$,
and $\Theta = a_0^2-2a_1a_{-1}$ is the singlet pair destruction operator. 
The states with the same spin but $M=0$ are obtained by applying the $S_-$ operator $2k$ times, yielding $
|2k, 0 \rangle = S_-^{2k} |2k, 2k \rangle/\sqrt{(4k)!}
$
with $S_- = \sqrt{2} \, (a_0^\dagger a_1 + a_{-1}^\dagger a_0)$.  
Since $[\Theta, S_-]=0$, we can permute the order of these two operators, and we are lead to consider $(a_0^\dagger a_1 + a_{-1}^\dagger a_0)^{2k} a_1^{\dagger 2k} |\emptyset\rangle = \sum_{j=0}^{2k} ((a_0^{\dagger 2k-j} a_0^j )) \,a_{-1}^{\dagger j} a_1^{2k-j} a_1^{\dagger 2k}|\emptyset\rangle$, where we define $((b^ic^j))$ as the sum of all different orders in which the $i+j$ operators can be arranged.  We can show by induction that $(( a_0^{\dagger 2k-j} \, a_0^{j} ))|\emptyset \rangle = {(2k)!\over (2k-2j)!\, 2^j\, j!} \, a_0^{\dagger 2(k-j)}|\emptyset \rangle$ and binomially expanding $\Theta^{\dagger n-k}$,  
$\langle p |2k, 0\rangle = A(k,n) p!\,\sqrt{(2n-2p)!}{(-2)^p\over (-4)^k} \sum_{s={\rm max}(0,k-p)}^{{\rm min}(k, n-p)} $ ${(-4)^s(2k)!^2\,(n-k)!\over (2s)!\, (k-s)!^2\, (n-p-s)!\, (p-k+s)!}$.  For $p=0$ using Stirling's approximation we get (\ref{02kapp}).

To verify eq. (\ref{p2k0}) we use that $(N+s)! \approx N! N^s$ for $s \ll N$, and thus obtain $\langle p |2k, 0\rangle\approx \sqrt{(4k_+1)(2n-2p)!\over 2n (2n)!} \, {(-2)^p\, (2k)!^2\, n! \over (-4)^k \,(n-p)! \, k!^2}\, \left ( {p\over n} \right )^k _2F_1(-k,-k;{1\over 2};1-{n\over p})$ where $_2F_1$ is the hypergeometric function.  Using now that 
~\cite{Erdelyi}
$_2F_1(\alpha, \beta;\gamma;z) = (1-z)^{-\alpha} \,_2F_1(\alpha, \gamma-\beta;\gamma;{z\over z-1})$
and that~\cite{Abramowitz} 
$
_2F_1(-k, k+1+\alpha+\beta;\alpha+1;x) =  {k! \, \Gamma(\alpha +1) \over \Gamma(\alpha+1+k)}\, P_k^{(\alpha, \beta)} (1-2x)
$
where $P_k^{(\alpha, \beta)} (z)$ is a Jacobi polynomial, and finally using that for $k \gg 1$ (although still $k \ll n, p$)
$P_k^{(-1/2,0)}(\cos \phi) \approx {1\over \sqrt{\pi \, k\, \cos(\phi/2)}}\, \cos\left [ (k+{1\over 4}) \phi \right ]
$
as well as the Stirling approximation 
we arrive at (\ref{p2k0}).

This work is supported by  NSF Grant DMR-0426149 and PHY-0555576.


\begin{thebibliography}{99}
\bibitem{Hospinor} T.-L. Ho, Phys. Rev. Lett. {\bf 81}, 742 (1998).
\bibitem{experiments} M.-S. Chang, et.al. Phys. Rev. Lett. {\bf 92} 140403 (2004); 
H. Schmaljohann et.al. Phys. Rev. Lett. {\bf 92} 040402 (2004). 
\bibitem{Pu}H. Pu et al., Phys. Rev. A {\bf 60}, 1463 (1999).
\bibitem{Law} C. K. Law, et al., Phys. Rev. Lett. {\bf 81}, 5257 (1998).
\bibitem{note on c} The constant $c$ is calculated as
$c = g \int d^3r |\varphi (r)|^4$, $g = \hbar (a_2-a_0) /M$, where $\varphi(r)$ is the spatial dependence of the wavefunction which is the same for all spin components. The dynamics is independent of the sign of $c$.
\bibitem{Berry}  M. V. Berry, J. Phys. A: Math. Gen. {\bf 29} 6617 (1996); Frank Grossmann et al., J. Phys. A: Math. Gen. {\bf 30} L277 (1997).
\bibitem{not valid for theta = 0} We note that this formula is in principle not applicable near $\theta = 0, \pi$.  This restriction is, however, not important for the interpretation of the quantum dynamics.
\bibitem{book} K. Chandrasekaran, {\em Elliptic Functions}
(Springer-Verlag, Berlin, 1980).
%
\bibitem{Erdelyi}
I. S. Gradshteyn and I. M. Ryzhik,
{\emph Table of integrals, series, and products, Corrected and
enlarged edition},
(Academic Press, New York, 1980).
\bibitem{Abramowitz}
Abramowitz and Stegun, formula 15.4.6.
\bibitem{longer paper} R. B. Diener (in preparation).
\bibitem{HoYip} T.-L. Ho and S.-K. Yip, Phys. Rev. Lett. {\bf 84}, 4031 (2000).
\end{thebibliography}
\end{document}